\documentclass[prc,preprintnumbers,showpacs,
showkeys,superscriptaddress,showpacs,nofootinbib,
byrevtex,fleqn]{revtex4-1}
\usepackage{amsmath,amsfonts,amssymb,amscd,amsxtra,amsthm}
\usepackage{graphicx,wrapfig,color}
\usepackage{bm}

\begin{document}
\preprint{INHA-NTG-07/2012}

\title{In-medium modified $\pi$-$\rho$-$\omega$ mesonic Lagrangian and 
  properties of nuclear matter} 

\author{Ju-Hyun Jung}
\email{juhyun@inha.edu}
\affiliation{Department of Physics, Inha University,
Incheon 402-751, Republic of Korea}

\author{Ulugbek T. Yakhshiev}
\email{yakhshiev@inha.ac.kr}
\affiliation{Department of Physics, Inha University,
Incheon 402-751, Republic of Korea}

\author{Hyun-Chul Kim}
\email{hchkim@inha.ac.kr}
\affiliation{Department of Physics, Inha University,
Incheon 402-751, Republic of Korea}

\date{November, 2012}
\begin{abstract}
We investigate the bulk properties of symmetric nuclear matter
within the framework of an in-medium modified chiral solitonic model 
with $\pi$, $\rho$ and $\omega$ mesons. We consider the modification
of meson degrees of freedom in nuclear matter, based on phenomenology
of pion-nucleus scattering and the empirical nuclear mass formula. 
We discuss the results of the density dependence of the volume term in
the mass formula and the incompressibility of symmetric nuclear
matter, comparing them with relativistic mean-field models. The mass
dropping of the $\rho$ meson in nuclear matter is also obtained and
discussed. 
\end{abstract}


\maketitle

\textbf{1.}
Understanding the modification of hadron properties in 
a hot and dense medium is one
of the most important issues, since it is deeply related to the
restoration of chiral symmetry of
quantum chromodynamics (QCD)~\cite{Birse:1994cz,Brown:1995qt}. In
particular, it is of great importance to understand the change of
hadron masses in medium, since much of them is generated by the
spontaneous breakdown of chiral symmetry (SB$\chi$S) of QCD. The quark
condensate is regarded as an order parameter of SB$\chi$S and is known
to decrease as temperature and density increase. Eventually chiral
symmetry will be restored at certain critical temperature and
density. Thus, it implies that the masses of hadrons are expected to
be modified in medium. 

The mass dropping of the vector mesons in dense matter has been
extensively studied theoretically as well as experimentally (see a
recent review~\cite{Hayano:2008vn} and references therein). In the
early 1990s, Brown and Rho investigated the mass dropping of hadrons
in dense medium, based on the effective chiral Lagrangian with scale
invariance, and suggested the in-medium scaling law, also known as the 
BR scaling~\cite{Brown:1991kk}. 
According to the proposed scaling law, the masses of the vector mesons 
are expected to drop approximately $20\,\%$ at the normal nuclear
matter density. Hatsuda and Lee drew a similar conclusion within the
QCD sum rules~\cite{Hatsuda:1991ez}. A most recent reanalysis of the
QCD sum rules suggests about $15\,\%$ dropping of the $\rho$ meson
mass at the normal nuclear matter density~\cite{Kwon:2010fw}. The
medium modification of the $\rho$ and $\omega$ mesons in cold nuclear
matter was experimentally reported only very recently. The KEK-PS E325 
experiment measured the invariant mass spectra of $\rho,\,\omega\to
e^+e^-$ decay modes in proton-induced nuclear
reactions~\cite{Naruki:2005kd} and announced that the masses of the
$\rho$ and $\omega$ mesons decrease by about $9\,\%$ at the normal
nuclear matter density. On the contrary, the CLAS experiment found
almost no evidence of the mass dropping of the $\rho$
meson~\cite{Nasseripour:2007aa, Wood:2008ee}. Instead, the CLAS
experiment observed the broadening of the $\rho$ meson decay width
that is in line with the predictions by hadronic
models~\cite{Peters:1997va, Cabrera:2000dx}.     

The nucleon in medium also undergoes modification and eventually its
structure gets altered. A typical example is the well known nuclear
EMC effect~\cite{Aubert:1983xm}. A recent experiment of the ratio of
the electric and magnetic form factors of the proton in the
${}^4\mathrm{He}(\vec{e},\,e'\vec{p}) {}^3\mathrm{H}$ reaction
indicates that the electromagnetic (EM) form factors are changed in
nuclei~\cite{Strauch:2002wu}. The modification of the nucleon mass in
nuclear matter has been studied within various nuclear 
models such as the Walecka model (see a review~\cite{Serot:1984ey},
and references therein), the quark-meson coupling (QMC)
model (see a review~\cite{Saito:2005rv}, and references therein),
the soliton-bag model~\cite{Jandel:1983gz}, 
the Skyrme model~\cite{Rakhimov:1996vq,Yakhshiev:2010kf} and so on. 
However, there is no consensus about how much its mass decreases so
far. Depending on models, the effective nucleon mass ($M_N^*$) at the
normal nuclear matter density is known to be dropped in the range of 
$500\,\mathrm{MeV}\lesssim M_N^*\lesssim 930\,\mathrm{MeV} $.  On
the other hand, Celenza et al.~\cite{Celenza:1984ew} proposed that the
size of the nucleon increased in nuclear medium in order to explain
the EMC effect.     
 
In this Letter, we investigate the bulk properties of symmetric
nuclear matter within the framework of a chiral soliton model based on
the effective Lagrangian with the $\pi$, $\rho$, and $\omega$
meson degrees of freedom. This chiral solitonic approach has a certain
virtue that the mesonic sector governs the baryonic sector, so that
the medium modification of the mesons and nucleon can be treated on
the same footing. The chiral soliton model with vector mesons in free 
space~\cite{Fujiwara:1984,Igarashi:1985,Meissner:1986js, 
Meissner:1986ka,Meissner:1988PhR} has
been developed in 1980s such that the $\rho$ and $\omega$ mesons
stabilize the skyrmion in place of the original stabilizing Skyrme
term~\cite{Skyrme:1961vq,Skyrme:1962vh,Adkins:1983ya}. The Skyrme term
is known to be related to the massive vector
mesons~\cite{Ecker:1989yg}. The chiral soliton model with vector
mesons was shown to describe well the static properties of the nucleon
in free space~(see the review~\cite{Meissner:1988PhR} and references
therein). In the meanwhile, the original Skyrme model was extended to
finite density~\cite{Rakhimov:1996vq}, the
pion properties being modified in a dense medium based on the experimental
data of pion-nucleus scattering. In Ref.~\cite{Yakhshiev:2010kf}, the
model was further elaborated by revising the stabilizing Skyrme term
in nuclear matter. The medium-modified Skyrme model improved   
the selfconsistency of the approach and could describe the basic
properties of nuclear matter. In the present work, we propose an
in-medium chiral soliton model with vector mesons. The model has an
advantage that it describes the meson and nucleon properties in free
space and in nuclear matter, and the bulk properties of nuclear matter
on an equal footing. In particular, the model shows one possible way
as to how recent issues of the mass dropping of the vector mesons can
be related to the properties of nuclear matter.

\vspace{0.5cm}
\textbf{2.}
Our starting point is the effective chiral Lagrangian 
with the $\pi$, $\rho$, and $\omega$ meson degrees of 
freedom~\cite{Meissner:1986js,Meissner:1986ka}, where the nucleon 
arises as a topological solution. We extend the model by
introducing the medium functionals that encode the influence of
the surrounding nuclear environment to the nucleon structure and 
the corresponding properties in line with the previous 
works~\cite{Rakhimov:1996vq,Yakhshiev:2010kf}, where the ``outer
shell''~\cite{Rakhimov:1996vq} and ``inner core''~\cite{Yakhshiev:2010kf}
modifications of the nucleon in nuclear matter have been
considered. The corresponding in-medium modified Lagrangian has the
following form\footnote{Hereafter, the asterisk indicates explicit
  medium modifications.} 
\begin{equation}
 \label{eq:1}
\mathcal{L}^{*} \;=\; \mathcal{L}_\pi^{*}
+\mathcal{L}_{V}^*+\mathcal{L}_{\pi V}^*+\mathcal{L}_{WZ}^*\,,  
\end{equation}
where the kinetic and mass terms are written as 
\begin{eqnarray}
\mathcal{L}_\pi^{*} & = & \frac{f_{\pi}^{2}}{4}\,
\mbox{Tr}\left(\partial_{0}U\partial_{0}U^{\dagger}\right)
-\alpha_{p}(\rho)\,\frac{f_{\pi}^{2}}{4}
\mbox{Tr}\left(\partial_{i}U\partial_{i}U^{\dagger}\right) \,+\,
\alpha_{s}(\rho)\,\frac{f_{\pi}^{2}m_{\pi}^{2}}{4}\,
\mbox{Tr}\left(U+U^{\dagger}-2\right),\cr
\mathcal{L}_{V}^* & = &
-\frac{1}{2\left(g_{V}^*\right)^2}\,\mbox{Tr}\left\{\partial_\mu
  V_\nu^*-\partial_\nu V_\mu^*-i[V_\mu^*,V_\nu^*]\right\}^2,
\end{eqnarray}
where $g_V^*$ stands for the generic coupling constant for a
vector meson. The vector fields are defined as
\begin{equation}
V_{\mu}^{(\rho)*} \;=\; \frac{g_\rho^*}{2}\,\tau^{a}\rho_{\mu}^{a}\,,
\qquad V_{\mu}^{(\omega)*} \;=\; \frac{g_\omega^*}{2}\,\omega_{\mu}, 
\end{equation}
and the interaction part has the following form 
\begin{equation}
  \label{eq:2}
\mathcal{L}_{\pi V}^* \;=\; 2f_\pi^2
\,\mbox{Tr}\left(J_{\mu}-V_{\mu}^{(\rho)*}-V_{\mu}^{(\omega)*}\right)^{2}.
\end{equation}
The vector current with the $\pi$ field is defined as 
\begin{equation}
  \label{eq:3}
J_{\mu} \;=\; \frac{1}{2i}\,\frac{\partial_{\mu}UU^{\dagger}+
\partial_{\mu}U^{\dagger}U}{\det\left(1+U\right)}.
\end{equation}
The last term in Eq.(\ref{eq:1}) denotes the Wess-Zumino term 
\begin{equation}
  \label{eq:4}
\mathcal{L}_{WZ}^* \;=\; \frac{3}{2}\,g_\omega^*
\omega_{\mu}B^\mu\,  
\end{equation}
which is related to the topological current 
\begin{equation}
  \label{eq:5}
B^\mu \;=\; \frac{\epsilon^{\mu\nu\alpha\beta}}
{24\pi^{2}}\,\mbox{Tr}\left\{ \left(U^{\dagger}\partial_{\nu}U\right)
\left(U^{\dagger}\partial_{\alpha}U\right)
\left(U^{\dagger}\partial_{\beta}U\right)\right\}.  
\end{equation}
The temporal component of Eq.(\ref{eq:5}) is identified as the density 
of the single baryon located in free space or in nuclear matter,
satisfying $\int d^3 rB^0=1$.

The pion fields are parametrized as a form of the SU(2) matrix
$U=\exp(i\bm{\tau}\cdot\bm{\pi}/f_\pi)$, while $\rho_\mu^a$ and
$\omega_\mu$ represent the isovector-vector and the isoscalar-vector
fields. We have introduced the additional medium functionals
  $\zeta_\rho(\rho)$ and $\zeta_\omega(\rho)$ such that
the effective in-medium couplings are changed in nuclear matter as
follows: $g_\rho^*=g\zeta_\rho(\rho)^{1/2}$ and   
$g_\omega^*=g\zeta_\omega(\rho)^{1/2}$.
The input parameters $f_{\pi}$, $m_\pi$ and $g$ stand
for the pion decay constant, the pion mass and the $\rho\pi\pi$
coupling constant $g=g_{\rho\pi\pi}$ in free space, respectively. 
The medium functionals $\alpha_s(\rho)$ and $\alpha_p(\rho)$ in
Eq.~~(\ref{eq:1}) are related to the phenomenology of low-energy
pion-nucleus scattering~\cite{Rakhimov:1996vq,Ericson}. 
The $\zeta_\rho(\rho)$ and  $\zeta_\omega(\rho)$ are considered in the 
kinetic term of the vector mesons and interaction parts
phenomenologically. Later, we will show that they  
will influence the inner core of the nucleon. In fact,
it is similar to the modification of the skyrme parameter 
$e\rightarrow e^*=e\zeta^{1/2}$ as done in
Ref.~\cite{Yakhshiev:2010kf}. The modified  
Skyrme term presented in Ref.~\cite{Yakhshiev:2010kf} stabilizes 
not only the solitons of the model but also homogeneous nuclear
matter. Similarly, the modified $\rho$ and $\omega$ 
mesons stabilize the solitons and reproduce the properties of
symmetric nuclear matter.  
We want to mention that the masses of the $\rho$ and $\omega$ mesons 
are degenerate in the original Skyrme model with the vector
mesons~\cite{Meissner:1986js,Meissner:1986ka}   
and they are related to the input parameters by the KSRF
relation~\cite{KSRF}: $m_{\rho}^{2} = m_{\omega}^{2} =  2 f_{\pi}^{2}
g^{2}$ in free space, i.e. in the case of $\alpha_p(0)=1$,
$\alpha_s(0)=1$ and $\zeta_\rho(0)=\zeta_\omega(0)=1$. However, we
consider two different models in the present work. First, we keep the 
masses of the $\rho$ and $\omega$ mesons degenerate, assuming that the
KSRF relation holds also in nuclear matter
\begin{eqnarray}
2 f_{\pi}^{2} g^{2}\zeta=m_{V}^{*\,2}\equiv m_{\rho}^{*\,2} =
m_{\omega}^{*\,2}, \qquad \zeta_\rho=\zeta_\omega\equiv\zeta,
\end{eqnarray} 
but the coupling constant in the Lagrangian will undergo a change in
nuclear medium as $g\rightarrow g_V^*=g\zeta^{1/2}$. We call it
\textit{Model I}. On the other hand, we will also assume that the KSRF 
relation in nuclear matter is valid only for the $\rho$ meson with the
$\omega$ meson intact, i.e.  
\begin{eqnarray}
  2 f_{\pi}^{2} g^{2}\zeta= m_{\rho}^{*\,2} \ne
  m_{\omega}^{*\,2}=m_\omega^2\,\qquad \zeta_\rho=\zeta,\qquad
  \zeta_\omega=1.  
\end{eqnarray}
This will be called \textit{Model II}.  
In doing so, we can understand how  
the $\rho$ and $\omega$ mesons play roles differently in describing
the properties of nuclear matter. The decay constant and mass of the
pion in free space are fixed by the experimental data,
i.e. $f_\pi=93$~MeV and $m_\pi=135$~MeV (the neutral pion mass). The
coupling constant is chosen as $g=5.85$, which is close 
to the empirical value $g_{\rho\pi\pi}=6.11$ that is determined from
the width of the $\gamma\rightarrow 2\pi$ decay. The vector
meson mass in free space is taken to be $m_V=m_\rho=m_\omega=770$~MeV.  

The medium functionals are given
as~\cite{Rakhimov:1996vq,Yakhshiev:2010kf,Ericson}
\begin{equation}
  \label{eq:9}
\alpha_p \;=\; 1-\frac{4\pi c_0 \rho /\eta}
{1 +g_0' 4\pi c_0 \rho/\eta},\;\;\;\;\;\;
\alpha_s \;=\; 1 - \frac{4\pi \eta b_0 \rho}{m_\pi^2},\;\;\;\;\;\;
\zeta \;=\; \exp\left\{-\frac{\gamma_{\rm num}\rho}{1+\gamma_{\rm 
      den}\rho}\right\},  
\end{equation}
where $\eta=1+m_\pi/m_N=1.14$ is a kinematical factor. The 
empirical parameters $b_0=-0.024\,m_\pi^{-1}$ and
$c_0=0.09\,m_\pi^{-3}$ are in general consistent with the analysis of
pionic atoms and the data of low-energy pion-nucleus scattering, and
$g_0'=0.7$ is an empirical value of the correlation
parameter~\cite{Ericson}. 

Most solitonic models yield larger values of the nucleon mass 
and the present model gives $m_N=1526$~MeV in free
space~\cite{Meissner:1988PhR}, whereas the mass splittings are
correctly reproduced. Thus, instead of the absolute value
of the nucleon mass, we consider the following parameter directly
related to the coefficient of the volume term in the Weizs\"acker mass 
formula for nuclei as done in Ref.~\cite{Yakhshiev:2010kf}.
One can minimize the mass functional derived from the Lagrangian
and obtain the effective nucleon mass $m_N^*$ in nuclear matter for  
a given set of parameters $\gamma_{\rm num}$ and $\gamma_{\rm den}$.
Then, the binding energy per nucleon can be defined as
\begin{equation}
m_N^*-m_N\equiv \varepsilon=-a_V(\rho)+\cdots,
\label{BinEn}
\end{equation}
where $a_V(\rho)$ is the density-dependent coefficient for the volume
term and the dots represent the surface, Coulomb, symmetry energy and
pairing terms. Since we consider an infinite isospin symmetric matter,
we concentrate only on the volume term and ignore all other terms in
the nuclear mass formula. The density dependence of $a_V(\rho)$ is 
obvious, because it describes the binding energy per nucleon in
symmetric nuclear matter. We rewrite Eq.(\ref{BinEn}) as
\begin{equation}
\frac{m_N^*}{m_N}= \frac{m_N-a_V(\rho_0)}{m_N}
\simeq \frac{m_N^{\rm (exp)}-a_V^{\rm (exp)}(\rho_0)}{m_N^{\rm (exp)}},
\label{BErel}
\end{equation}
so that the relation will be satisfied at the saturation point of
nuclear matter, $\rho=\rho_0$. The parameters $\gamma_{\mathrm{num}}$
and $\gamma_{\mathrm{den}}$ are fitted to reproduce the correct
saturation point. Note that experimentally $m_N^{\rm (exp)}=939$~MeV
and $a_V^{\rm (exp)}(\rho_0)=16$~MeV. However, almost all chiral
soliton models produce the larger values of the nucleon mass. 
In the Skyrme model with vector mesons, the nucleon mass typically
turns out to be around 1500 MeV, while the $\Delta$-nucleon mass
difference in free space $\Delta M = M_\Delta - M_N$ is obtained to be
similar to the experimental data. Since the nucleon mass is
overestimated by about $40\,\%$, it is better to use $a_V(\rho_0)=26$
MeV phenomenologically, which is about $40\,\%$ larger than the  
typical value $a_V^{\rm (exp)}(\rho_0)=16$~MeV. For example, 
in Ref.~\cite{Yakhshiev:2010kf}, the experimental value of the nucleon
mass, i.e. $m_N =939$ MeV was shown to reproduce $a_V=16$ MeV 
correctly. 

\begin{figure}[ht]
\centerline{\includegraphics[width=6cm]{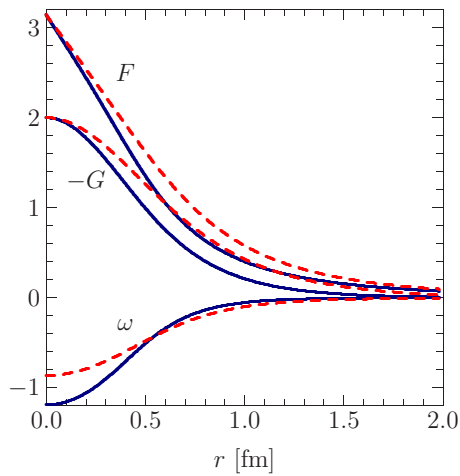}}
\caption{(Color online) Meson profile functions in free space (solid
  curves) and and at the normal nuclear matter density
  $\rho_0=0.15~{\rm fm}^{-3}$ (dashed ones). The results are those
  from the Model I.    
 }
\label{fig:n1}
\end{figure}

As mentioned already, we are interested in \textit{symmetric}
nuclear matter. Since the nuclear density $\rho$ is constant, one can
use the spherically symmetric approximation for the meson profiles,
i.e. the hedgehog Ans\"atze  
\begin{equation}
  \label{eq:12}
U \;=\; \exp\left\{i\bm{\tau}\cdot \hat{\bm{n}} F(r)\right\},\;\;\;\;
\hat{\bm{n}} \;=\; \frac{\bm{r}}{r},\;\;\;\;
\rho_i^a \;=\; \varepsilon_{iak}n_k\,\frac{G(r)}{r L^*},\;\;\;\;
\omega_\mu \;=\; \omega(r)\delta_{\mu0},  
\end{equation}
where the scale factor $L^*=g_\rho^*$ is required to have the same
boundary conditions for the medium-modified profile function $G$ as in
free space, i.e. $G(0)=-2$ (see Fig.~\ref{fig:n1}).  

Using the hedgehog Ans\"atze, one can derive the static energy
functional from the Lagrangian, which is identified as the classical
skyrmion mass 
\begin{eqnarray}
M_H^*&=& E^*[F(r),G(r),w(r)] \cr
&=&-\int\mathcal{L}^*\mathrm{d}^{3}r \cr
&=& 4\pi\int_0^\infty {\rm
  d}r\left\{\alpha_{p}\,\frac{f_{\pi}^{2}}{2}\left(r^2F^{\prime\,2} 
 +{2\sin^{2}F}\right)+\,{2f_{\pi}^{2}}\left(G+1-\cos F\right) ^{2}
 +\alpha_{s}r^2f_{\pi}^{2}m_{\pi}^{2}\left(1-\cos F\right)\right.\cr
 && \left.\hspace{1.8cm} +\,\frac{1}{g_\rho^{*2}}\left[G^{\prime\,2} +
     \frac{G^2(G+2)^2}{2r^2} \right]
-\,r^2\left(f_\pi^2g_\omega^{*2}\omega^2+\frac12\,\omega^{\prime\,2}\right) +
   \frac{3g_{\omega}^*}{4\pi^2}\,\omega F'\sin^2F \right\},
\end{eqnarray}
where $f'=\partial f/\partial r$, generically. 
Minimizing the  classical soliton mass can be achieved by solving the
equations of motion, which are given as the coupled nonlinear
differential equations 
\begin{eqnarray}
F^{\prime\prime}&=&-\frac{2}{r}\,
F'+\frac{1}{\alpha_pr^2}\left[4(G+1)\sin F-\alpha_p\sin
  2F\right]  + \frac{\alpha_s}{\alpha_p}\, m_\pi^2 \sin
F-\frac{3g_\omega^{*}}{4\pi^2\alpha_pf_\pi^2}\,
\frac{\omega'}{r^2}\sin^2F\,,\cr  
G^{\prime\prime}&=&2g_\rho^{*2} f_\pi^2\left[G+2\cos\frac{F}2\right] +
\frac{G(G+1)(G+2)}{r^2} 
\nonumber\,,
\qquad\\
\omega^{\prime\prime}&=&-\frac{2}{r}\,\omega' + 2f_\pi^2
g_\omega^{*2}\omega-\frac{3g_\omega^*}{4\pi^2r^2}F'\sin^2F\, 
\label{statEQ}
\end{eqnarray}
with the boundary conditions
\begin{equation}
  \label{eq:15}
F(0) \;=\; \pi,\;\;\;\;\; F(\infty)\;=\; 0,\;\;\;\;\;
G(0)=-2,\;\;\;\;\; G(\infty)=0, \;\;\;\;\;
\omega'(0) \;=\; 0,\;\;\;\;\; \omega(\infty)=0\,.
\end{equation}

The collective quantization brings out the following relations
\begin{eqnarray}
U(\bm{r},\,t) &=& A(t)U(\bm{r})A^+(t),\;\;\;\; 
\omega_i(\vec r,t) \;=\; \frac{\phi(r)}{r}\, \left(\bm{K}\times
  \bm{n}\right)_i, \cr 
\bm{\tau}\cdot\bm{\rho_0} (\bm{r},\,t) &=& \frac{2}{g_\rho^*} A(t)
\bm{\tau}\cdot[\bm{K} \xi_1(r) + \bm{n}\cdot \left(\bm{K} \times
  \bm{n}\right) \xi_2(r)] A^+(t),\cr
\bm{\tau}\cdot\bm{\rho}_i(\bm{r},\,t) &=& A(t) \bm{\tau} \cdot
\bm{\rho}_i(\bm{r}) A^+(t),
\end{eqnarray}
where $2\bm{K}$ denotes the angular velocity of the soliton with   
the relation $i\bm{\tau}\cdot \bm{K} = A^+\dot{A}$. This leads to the
time-dependent collective Hamiltonian
\begin{equation}
H^*(t) \;=\; M_H^*(F,\rho,\omega) -
\Lambda^*(F,\rho,\omega,\phi,\xi_1,\xi_2)\mbox{Tr}(\dot A\dot A^+),   
\end{equation}
where $\Lambda^*$ denotes the in-medium modified moment of inertia of
the rotating skyrmion 
\begin{eqnarray}
\Lambda^*&=& \frac{g_\omega^*}{2\pi}\,\phi
F'\sin^2F+\frac23f_\pi^2r^2 
\left(\sin^2\!F\!+\!8\sin^4\frac{F}{2}-8\xi_1\sin^2\frac{F}{2}+
2\xi_1^2\!+\!(\xi_1\!+\!\xi_2)^2\!\right)\cr
&& +\; \frac{1}{3g_\rho^{*2}}\left[4G^2(\xi_1^2 + \xi_1\xi_2 - 2\xi_1 -
  \xi_2 + 1) + 2(G^2+2G+2)\xi_2^2+r^2(3\xi_1^{\prime2} +
  \xi_2^{\prime2} + 2\xi_1^{\prime}\xi_2^{\prime})\right]\cr
&&-\; \frac{1}{6}\left(\phi^{\prime2}+\frac2{r^2}\phi^2\right)
-\frac{1}3\,f_\pi^2g_\omega^{*2}\phi^2.
\end{eqnarray}
In the large $N_c$ expansion, one extremizes the moment of inertia
and gets the coupled nonlinear differential equations for the next-order 
profile functions $\xi_1$, $\xi_2$, $\phi$
in the presence of the leading-order profile functions $F$, $G$ and
$\omega$ 
\begin{eqnarray}
\xi_{1}'' & = & 2f_{\pi}^{2}g_\rho^{*2} \left(\xi_{1}-1+\cos
  F\right)-\frac{2\xi_{1}'}{r} +
\frac{G^{2}\left(\xi_{1}-1\right)+2(G+1)\xi_{2}}{r^{2}},\cr
\xi_{2}'' & = & 2f_{\pi}^{2}g_\rho^{*2}\left(\xi_{2}+1-\cos F\right)
-\frac{2\xi_{2}'}{r} + \frac{G^{2}\left(\xi_{1}-1\right) + 
2\left(G^2+3G+3\right)\xi_{2}}{r^{2}},\cr
\phi'' & = & 2f_{\pi}^{2}g_\omega^{*2}\phi - 
\frac{3g_\omega^*F'\sin^{2}F}{2\pi^{2}} + \frac{2\phi}{r^{2}}
\end{eqnarray}
with the boundary conditions
\begin{equation}
  \label{eq:20}
\phi(0) \;=\; \phi(\infty) \;=\; 0,\;\;\;\;
\xi_1'(0)\;=\; \xi_1(\infty) \;=\; 0,\;\;\;\;
\xi_2'(0)\;=\; \xi_2(\infty) \;=\; 0.  
\end{equation}
The boundary conditions for $\xi_1$ and $\xi_2$ satisfy the relation
$2\xi_1(0)+\xi_2(0)=2$ that remains unchanged in nuclear matter. 

Finally, the effective masses of the nucleon and the $\Delta$ isobar
are expressed in terms of the in-medium hedgehog mass $M_H^*$ and the
in-medium modified moment of inertia $\Lambda^*$ 
\begin{equation}
  \label{eq:21}
m_N^*=M_H^*+\frac{3}{8\Lambda^*},\;\;\;\;\;\;
m_\Delta^*=M_H^*+\frac{3}{15\Lambda^*}. 
\end{equation}

\begin{figure}[ht]
\centerline{\includegraphics[width=6cm]{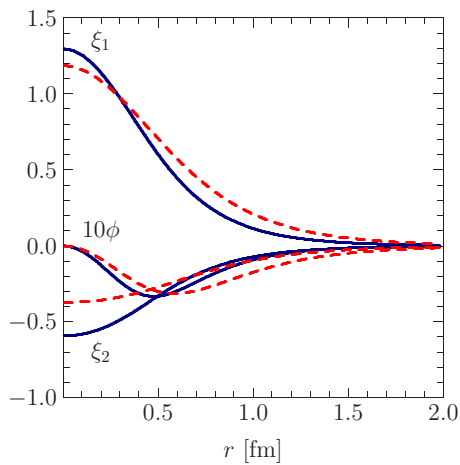}}
\caption{(Color online) The profile functions $\xi_1$, $\xi_2$, and 
  $10\,\phi$. The solid curves depict them in free space, whereas
  dashed ones represent them at the normal nuclear matter
  density $\rho_0=0.15~{\rm fm}^{-3}$. 
  The results are those from the Model I.
}
\label{fig:n2}
\end{figure}

\vspace{0.5cm}
\textbf{4.}
Figure~\ref{fig:n1} draws the solutions of the equations of
motion~(\ref{statEQ}) for Model I, i.e. the profile functions for 
the $\pi$, $\rho$, $\omega$ background fields. The solid curves depict
those in free space, whereas the dashed ones represent the modified
profile functions at the normal nuclear matter density
$\rho_0=0.15~{\rm fm}^{-3}$. One can clearly see that the profile
functions $F(r)$ and $G(r)$ are enhanced everywhere. On the contrary, the
$\omega$ one is suppressed up to around $r\approx0.6\,\mathrm{fm}$, it
starts to be strengthened. This is partly due to the fact that the
profile functions $F(r)$ and $G(r)$ are constrained to be $\pi$ and $2$ at
$r=0$ by the Dirichlet boundary conditions, whereas $\omega(r)$ has the
Cauchy boundary conditions. 
We also found that all the profile functions showed similar behaviors
in the case of Model II. 
In Fig.~\ref{fig:n2}, the profile functions $\xi_1$, $\xi_2$ and
$\phi$ for Model I are illustrated.
These three profile functions also show enhancement in larger
distances but are suppressed in smaller distances. The similar results
are obtained also in the case of Model II. 

\begin{figure}[ht]
\centerline{\includegraphics[width=6cm]{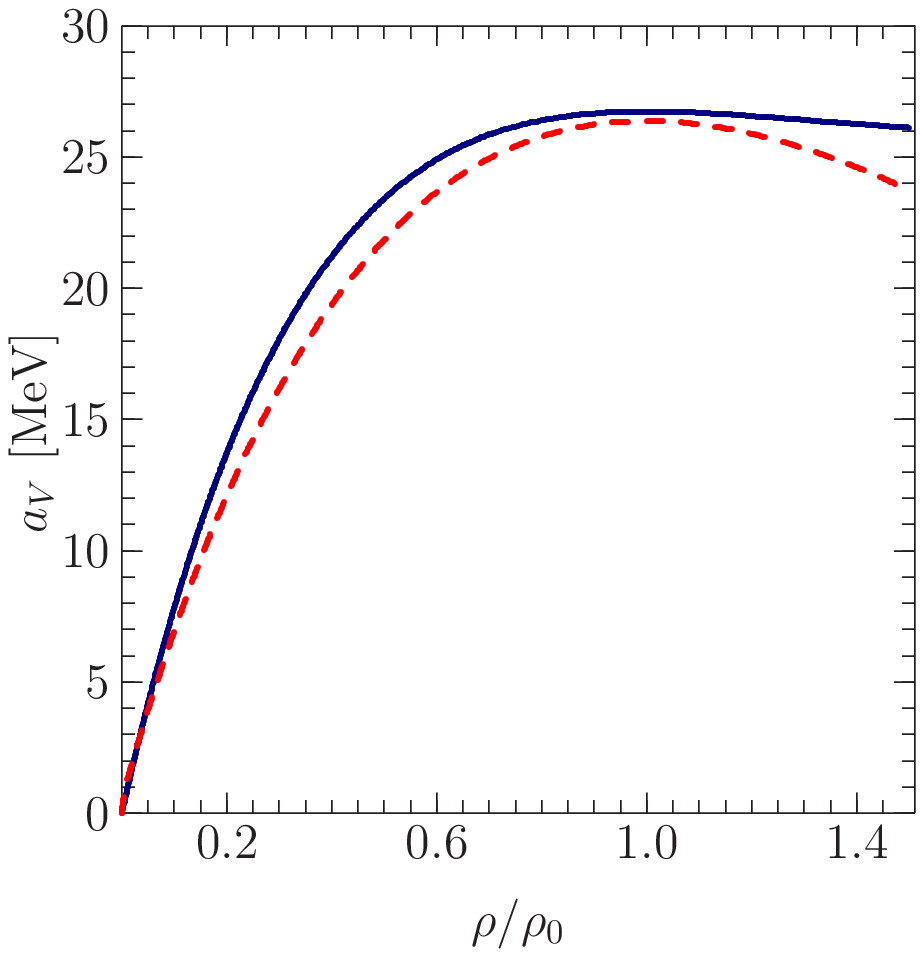}}
\caption{(Color online) The coefficient of the volume term in the
  binding energy formula~(\ref{BinEn}) $a_V$ as a function of 
  nuclear matter density normalized by the normal nuclear matter
  density, i.e. $\rho/\rho_0$. The solid curve draws the result of
  Model I while the dashed one represents that of Model II. 
}
\label{fig:n3}    
\end{figure}
Figure~\ref{fig:n3} shows the results of the coefficient of the volume 
term $a_V$ as a function of nuclear matter density normalized by
the normal nuclear matter density, i.e. $\rho/\rho_0$. The solid curve 
represents the result of Model I. In this case, $a_V$ 
increases as the density increases and then is already saturated to be
approximately $a_V\simeq 26$ MeV at about $\rho\sim 0.6\,\rho_0$ 
and remains almost constant up to the normal nuclear matter
density. Then, it decreases very slowly.  
It is very similar to those obtained within the in-medium chiral
soliton model, where the Skyrme term was used instead of those with 
the vector mesons~\cite{JKPS60-356}. This is due to the fact that 
the overall changes of the $\rho$ and $\omega$ mesons in nuclear
matter behave in almost the same way as the change of 
the Skyrme term as presented in Ref.~\cite{JKPS60-356}. On the other
hand, we find that $a_V$ has a clear minimum for the binding energy at
the normal nuclear matter density in the case of Model II, as shown in
the dashed curve in Fig.~\ref{fig:n3}. In Model II, the $\omega$ meson
remains intact, whereas the $\rho$ meson undergoes a change in nuclear
matter. It implies that a inner core part of the nucleon   
is less modified than its outer shell part. Thus, the result of Model
II, which is more favorable (see the results in Table~\ref{tab:1})
than that of Model I in the present approach, 
can be interpreted as follows: as the density increases, the
wavefunctions of nucleons in medium are forced to be overlapped but
the effects of the Pauli blocking resist this overlapping. It means
that while the outer shell part of a nucleon can be easily influenced,
the inner core part is less affected by its neighborhoods. 

One can make a Taylor expansion of the coefficient of the volume term
near the normal nuclear matter density with respect to $\rho$
\begin{eqnarray}
a_V(\rho)&=&a_V(\rho_0)+\frac{1}{2}\left(\frac{\partial^2
    a_V}{\partial \rho^2}\right)_{\rho=\rho_0} 
(\rho-\rho_0)^2 + \frac{1}{6}\left(\frac{\partial^3 a_V}{\partial
    \rho^3}\right)_{\rho=\rho_0} 
(\rho-\rho_0)^3+\cdots \cr
&\equiv&a_0+\frac{K_0}{18}
(\rho-\rho_0)^2+\frac{Q_0}{162}
(\rho-\rho_0)^3+\cdots,
\end{eqnarray}
where $a_0$, $K_0$, and $Q_0$ denote the value of $a_V$ at
$\rho=\rho_0$, the incompressibility of symmetric nuclear matter, and  
the coefficient of the third derivative, respectively. The coefficient
of the first derivative, which is defined as the pressure, vanishes at
the saturation point. In Table~\ref{tab:1} we list the results of
these three coefficients in comparison with other models. The
parameters $\gamma_{\mathrm{num}}$ and $\gamma_{\mathrm{den}}$ are
fitted to reproduce $a_V\approx26$ MeV in both Model I and II. 
\begin{table}[ht]
\vskip 0.2cm
\begin{tabular}{l|ccccccc}
\hline\hline
  & $\gamma_{\mathrm{num}}$ & $\gamma_{\mathrm{den}}$ &
  $\rho_0\,[\mathrm{fm}^{-3}]$&$a_0\,[\mathrm{MeV}]$ &
  $K_0\,[\mathrm{MeV}]$ & $Q_0\,[\mathrm{MeV}]$ & 
  $m_\rho^*/m_\rho$ \\  
\hline
Model I & 2.390 & 1.172 &0.150& 26.75 & 110 & $-1591$ & 0.70\\
Model II& 1.970 & 0.841 &0.150& 26.38 & 233 & $-1213$ & 0.72 \\
Hybrid~\cite{Piekarewicz:2008nh} & -- & -- & 0.148 &16.24 & 230 &\,\,
$-72$  & -- \\ 
FSU~\cite{ToddRutel:2005zz} & -- & -- & 0.148 & 16.30 & 230 &\, $-523$
& -- \\ 
NL3~\cite{Lalazissis:1996rd} & -- & -- & 0.148 & 16.24 & 272 &
\quad$204$ & -- \\ 
\hline\hline
\end{tabular}
\caption{The bulk properties of nuclear matter: $\rho_0$
  denotes the saturation point of nuclear matter density that
  corresponds to the normal nuclear matter density;
  $a_0$ represents the value of $a_V$ at the saturation point; $K_0$
  designates the compressibility of symmetric nuclear matter; $Q_0$ is 
  the coefficient of the third derivative at $\rho_0$; 
  $m_\rho^*/m_\rho$ is the ratio of the mass of the $\rho$ 
  meson at the normal nuclear matter density to that of the
  vacuum. The first and second rows list the corresponding results
  from Model I and Model II, respectively, while the other three rows
  are those from
  Refs.~\cite{Piekarewicz:2008nh,ToddRutel:2005zz,Lalazissis:1996rd}.
}
\label{tab:1}
\end{table}
The incompressibility of symmetric nuclear matter $K_0$ is an
essential quantity, because it shapes the dependence of the nuclear
binding energy on the density near the saturation point. It also means
that it shows how stiff or soft symmetric nuclear matter is. In the
fifth column of Table~\ref{tab:1}, the results of the present models
and other three different models are compared. Model I gives 110 MeV
while Model II yields 233 MeV, which implies that the nuclear binding
energy turns out to be stiffer in the case of Model II, that is, it
increases more rapidly with density than that of Model I does. The
result of Model II is comparable with the predictions of the other
models. Compared with the empirical value of the incompressibility
extracted from the experimental data of the giant monopole resonance
$K_0= 235\pm 14$ MeV~\cite{Youngblood:1999zza}, the present result of
Model II is in good agreement with the data.    
The sixth column of Table~\ref{tab:1} lists the predictions of $Q_0$
in comparison with those of the other models. The present results for
$Q_0$ are quite larger than those of 
Refs.~\cite{Piekarewicz:2008nh,ToddRutel:2005zz,Lalazissis:1996rd}. 
This is due to the fact that the medium functionals
are determined from pion-nucleus scattering phenomenology at 
subnormal nuclear matter densities. If $\rho$ is larger than 
$\rho_0$, the effective parameters of the medium functionals and, in
particular, the correlation parameter $g_0'$ may become
density-dependent.   

\begin{figure}[ht]
\centerline{\includegraphics[width=6cm]{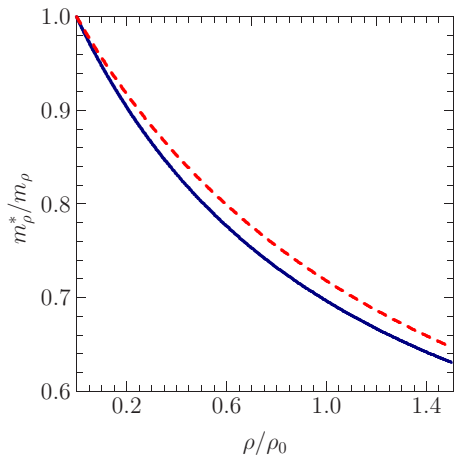}}
\caption{The ratio of the mass of the $\rho$ meson in
  nuclear medium to its value in free space as a function of
  normalized nuclear matter density $\rho/\rho_0$. The solid curve 
  represents the result of Model I, while the dashed one draws that of
  Model II.}
\label{fig:n4}       
\end{figure}

As mentioned earlier, an advantage of the present approach is that we
can treat the mesonic and baryonic sectors on the same footing.  
Since we fix the relevant parameters to reproduce the properties of
the nucleon and nuclear matter, we are able to find the $\rho$ meson
mass at a given density. In this way, we determine how the $\rho$
meson mass is dropping as the density increases. Figure~\ref{fig:n4}
depicts the mass dropping of the $\rho$ meson as a function of the
normalized nuclear matter density $\rho/\rho_0$.
In Model I, the masses of the $\rho$ and $\omega$ mesons are
degenerate and hence their changes in nuclear matter are identical. As 
shown in Fig.~\ref{fig:n4}, they decrease as the density
increases. However, their dependence on the nuclear matter density
does not seem to be linear, i.e. it is not satisfied with the linear
relation $m_\rho(\rho) = m_\rho(0) (1-C\rho/\rho_0)$ with constant $C$
but it seems to behave as a quadratic function of the density. At
the normal nuclear matter density 
$\rho_0$, the masses of the $\rho$ and $\omega$ mesons
are dropped by about $30\,\%$ in the case of Model I. 
Model II brings
about the $\rho$ meson mass dropping by about $27\,\%$, while that of
the $\omega$ meson is kept to be fixed.        
These results show a similar tendency to those obtained within the QCD
sum rules~\cite{Hatsuda:1991ez} and the generalized NJL 
model~\cite{Shakin:1993ta}, in which the mass of the $\rho$ meson is
approximately dropped by about $20\,\%$.

As discussed previously, chiral soliton models yield in general the
overestimated nucleon mass, so that we fix all the parameters to
produce $a_V=26$ MeV, not the experimental value
$a_V^{\mathrm{exp}}=16$ MeV. If the same argument applies to the mass
dropping of the $\rho$ meson, the $27-30\,\%$ mass dropping is more or
less comparable with those of the QCD sum rules and NJL model, though
it turns out to be larger than the recent measurement of the KEK-PS
E325 experiment. However, we want to emphasize that in the present
work we want to present a possible alternative way of explaining the
changes of the nucleon and the bulk properties of nuclear matter from
a different point of view. 
  
\vspace{0.5cm}
\textbf{5.}
In the present work, we aimed at investigating the medium modification
of the vector mesons and nucleon on an equal footing, and at studying
the bulk properties of symmetric nuclear matter, based on the
in-medium modified $\pi$-$\rho$-$\omega$ soliton model. We employed
the two different models, i.e. Model I in which we keep the
$\rho$ and $\omega$ mesons degenerate, and Model II, where we let
the $\rho$ meson undergo a change while the $\omega$ is kept to be
intact. The pionic sector was determined by using pion-nucleus
scattering data, while the vector mesons were assumed to be changed in
a similar manner to the Skyrme term in
Ref.~\cite{Yakhshiev:2010kf}. We first examined 
how the profile functions were changed in nuclear matter. Then, we
determined the coefficient of the volume term in the nuclear mass
formula. Since the nucleon mass from this model is overestimated, we
also considered reasonably the larger value of the volume-term
coefficient. We found that Model I produced the coefficient of the
volume term that became almost saturated around $a_V\simeq 26$ MeV. On the
other hand, Model II exhibited a clear minimum 
of the binding energy per nucleon at around $26$ MeV,
from which the coefficient of the volume term started to decrease. The
results from Model II were more favorable because of the following
reason: as the density increases, the outer pionic shell of a nucleon
is more influenced by its environment, while the change of its inner
core is less recognized. We also calculated the incompressibility of
nuclear matter, which is one of the most important bulk properties of
nuclear matter. We obtained $K_0=233$ MeV from Model II, which is in
agreement with relativistic mean-field models. Finally, we
determined the mass dropping of the $\rho$ meson that arose from the
fixing of the vector-meson parameters. It was found that the $\rho$
meson mass was dropped by about $27-30\,\%$. 

The present work sheds light on the bulk properties of nuclear matter
from a different point of view. While the soliton model with vector
mesons provide a simple framework in investigating the properties of
nuclear matter, it still brings about certain physical
implications: in this framework, the modification of
the vector mesons shows a certain correlation with the changes of the
nucleon and nuclear matter properties.
 Moreover, the present approach can be easily extended to 
asymmetric nuclear matter or neutron-rich nuclear matter, the effects
of isospin symmetry breaking being
considered~\cite{Meissner:2007ya,Meissner:2008mr,JKPS03}. Consequently,
it will make it possible to study the properties of neutron stars
and the structure of neutron-rich nuclei within the present
framework. The corresponding works are under way. 

\acknowledgments

This work is supported by Basic Science
Research Program through the National Research Foundation of Korea
(NRF) funded by the Korean government (MEST) (No. 2011-0023478
(J.-H.~J. and U.~Y.)) and (No. 2012004024 (H.-Ch.~K.)), respectively.


\begin{thebibliography}{99}
\bibitem{Birse:1994cz} 
  M.~C.~Birse,
J.\ Phys.\ G {\bf 20} (1994) 1537 [nucl-th/9406029].  

\bibitem{Brown:1995qt} 
  G.~E.~Brown and M.~Rho,
Phys.\ Rept.\  {\bf 269} (1996) 333  [hep-ph/9504250].  

\bibitem{Hayano:2008vn} 
  R.~S.~Hayano and T.~Hatsuda,
  Rev.\ Mod.\ Phys.\  {\bf 82} (2010) 2949  [arXiv:0812.1702
  [nucl-ex]].  

\bibitem{Brown:1991kk} 
  G.~E.~Brown and M.~Rho,
Phys.\ Rev.\ Lett.\  {\bf 66} (1991) 2720.  

\bibitem{Hatsuda:1991ez} 
  T.~Hatsuda and S.~H.~Lee,
  Phys.\ Rev.\ C {\bf 46} (1992) 34  

\bibitem{Kwon:2010fw} 
  Y.~Kwon, C.~Sasaki and W.~Weise,
Phys.\ Rev.\ C {\bf 81} (2010) 065203  [arXiv:1004.1059 [nucl-th]].

\bibitem{Naruki:2005kd} 
  M.~Naruki {\it et al.} [KEK-PS E325 Collaboration],
Phys.\ Rev.\ Lett.\  {\bf 96} (2006) 092301  [nucl-ex/0504016].

\bibitem{Nasseripour:2007aa} 
  R.~Nasseripour {\it et al.}  [CLAS Collaboration],
Phys.\ Rev.\ Lett.\  {\bf 99} (2007) 262302  [arXiv:0707.2324
[nucl-ex]].  

\bibitem{Wood:2008ee} 
  M.~H.~Wood {\it et al.}  [CLAS Collaboration],
Phys.\ Rev.\ C {\bf 78} (2008) 015201  [arXiv:0803.0492 [nucl-ex]].

\bibitem{Peters:1997va} 
  W.~Peters, M.~Post, H.~Lenske, S.~Leupold and U.~Mosel,
Nucl.\ Phys.\ A {\bf 632} (1998) 109  [nucl-th/9708004].  


\bibitem{Cabrera:2000dx} 
  D.~Cabrera, E.~Oset and M.~J.~Vicente Vacas,
Nucl.\ Phys.\ A {\bf 705} (2002) 90  [nucl-th/0011037].  

\bibitem{Aubert:1983xm}
  J.~J.~Aubert {\it et al.}  [European Muon Collaboration],
  Phys.\ Lett.\ B {\bf 123} (1983) 275.

\bibitem{Strauch:2002wu} 
  S.~Strauch {\it et al.}  [Jefferson Lab E93-049 Collaboration],
Phys.\ Rev.\ Lett.\  {\bf 91} (2003) 052301  [nucl-ex/0211022].

\bibitem{Serot:1984ey}
  B.~D.~Serot and J.~D.~Walecka,
  Adv.\ Nucl.\ Phys.\  {\bf 16} (1986) 1.  

\bibitem{Saito:2005rv}
  K.~Saito, K.~Tsushima and A.~W.~Thomas,
Prog.\ Part.\ Nucl.\ Phys.\  {\bf 58} (2007) 1  [hep-ph/0506314].

\bibitem{Jandel:1983gz}
  M.~J\"andel and G.~Peters,
Phys.\ Rev.\ D {\bf 30} (1984) 1117.  

\bibitem{Rakhimov:1996vq} 
  A.~Rakhimov, M.~M.~Musakhanov, F.~C.~Khanna and U.~T.~Yakhshiev,
  Phys.\ Rev.\ C {\bf 58} (1998) 1738  [nucl-th/9609049].  
\bibitem{Yakhshiev:2010kf}
  U.~Yakhshiev and H.~-Ch.~Kim,
  Phys.\ Rev.\ C {\bf 83} (2011) 038203  [arXiv:1009.2909 [hep-ph]].

\bibitem{Celenza:1984ew}
  L.~S.~Celenza, A.~Rosenthal and C.~M.~Shakin,
Phys.\ Rev.\ Lett.\  {\bf 53} (1984) 892; ibid., \textbf{53}
(1984) 1610.  

\bibitem{Fujiwara:1984}
  T.~Fujiwara, Y.~Igarashi, A.~Kobayashi, H.~Otsu, T.~Sato, S.~Sawada, 
  Prog. Theor. Phys. \textbf{74} (1985) 128.

\bibitem{Igarashi:1985}
  Y.~Igarashi, M.~Johmura, A.~Kobayashi, H.~Otsu, T.~Sato, S.~Sawada, 
  Nucl. Phys.  B \textbf{259} (1985) 721.

\bibitem{Meissner:1986js}
  Ulf-G.~Meissner, N.~Kaiser, W.~Weise,
  Nucl. Phys. A \textbf{466} (1987) 685.

\bibitem{Meissner:1986ka}
  Ulf-G.~Meissner, N.~Kaiser, A~Wirzba, W.~Weise, 
  Phys. Rev. Lett. \textbf{57} (1986) 1676.

\bibitem{Meissner:1988PhR}
  Ulf-G. Meissner:
  Phys. Rep. \textbf{161} (1988) 213.

\bibitem{Skyrme:1962vh}
  T.~H.~R.~Skyrme,
  Nucl.\ Phys.\  {\bf 31} (1962) 556.

\bibitem{Skyrme:1961vq}
  T.~H.~R.~Skyrme,
  Proc.\ Roy.\ Soc.\ Lond.\ A {\bf 260} (1961) 127.

\bibitem{Adkins:1983ya}
  G.S.~Adkins, C.R.~Nappi,  E.~Witten,
Nucl. Phys. B \textbf{228} (1983) 552.

\bibitem{Ecker:1989yg}
  G.~Ecker, J.~Gasser, H.~Leutwyler, A.~Pich and E.~de Rafael,
Phys.\ Lett.\ B {\bf 223} (1989) 425.  

\bibitem{Ericson}
  T.~E.~O.~Ericson and W.~Weise,
  ``\textit{Pions And Nuclei},''  (Clarendon, Oxford, UK, 1988).

\bibitem{KSRF} K. Kawarabayashi and M. Suzuki,
  Phys. Rev. Lett. \textbf{16} (1966) 255; Riazuddin and Fayyazuddin,
  Phys. Rev. \textbf{147} (1966) 1071. 

\bibitem{JKPS60-356}
  U.T. Yakhshiev,
  J. Korean Phys. Soc. \textbf{60} (2012) 356.

\bibitem{Piekarewicz:2008nh}
  J.~Piekarewicz and M.~Centelles,
      Phys.\ Rev.\ C {\bf 79} (2009) 054311
  [arXiv:0812.4499 [nucl-th]].

\bibitem{ToddRutel:2005zz}
  B.~G.~Todd-Rutel and J.~Piekarewicz,
  Phys.\ Rev.\ Lett.\  {\bf 95} (2005) 122501.  
\bibitem{Lalazissis:1996rd}
  G.~A.~Lalazissis, J.~Konig and P.~Ring,
  Phys.\ Rev.\ C {\bf 55} (1997) 540  [nucl-th/9607039].  

\bibitem{Youngblood:1999zza}
  D.~H.~Youngblood, H.~L.~Clark and Y.~-W.~Lui,
  Phys.\ Rev.\ Lett.\  {\bf 82} (1999) 691.  

\bibitem{Shakin:1993ta}
  C.M.~Shakin, W.D.~Sun, 
  Phys. Rev. C \textbf{49} (1994) 1185.

\bibitem{Meissner:2007ya}
  U.~-G.~Meissner, A.~M.~Rakhimov, A.~Wirzba and U.~T.~Yakhshiev,
  Eur.\ Phys.\ J.\ A {\bf 32} (2007) 299  [arXiv:0705.1603 [nucl-th]].
\bibitem{Meissner:2008mr}
  U.~-G.~Meissner, A.~M.~Rakhimov, A.~Wirzba and U.~T.~Yakhshiev,
  Eur.\ Phys.\ J.\ A {\bf 36} (2008) 37  [arXiv:0802.1455 [nucl-th]].
\bibitem{JKPS03}
    U.T. Yakhshiev, to appear in J. Korean Phys. Soc. (2013).
\end{thebibliography}
\end{document}